\documentclass[twocolumn,twocolappendix,appendixfloats]{aastex6}

\usepackage{subfigure}
\usepackage{threeparttable}
\usepackage{multirow}
\usepackage{remreset}

\usepackage{CJKutf8}

\collaborationName{Friends of AASTeX}
\shorttitle{line-locked NALs within BALs}
\shortauthors{Lu \& Lin}
\begin{document}


\title{Narrow absorption lines complex II: probing the line-locking signatures within trough-like broad absorption line}



\author{
\begin{CJK*}{UTF8}{gbsn}
Wei-Jian Lu (陆伟坚)\altaffilmark{1} and Ying-Ru Lin (林樱如)\altaffilmark{2}
\end{CJK*}
}
\affil{School of Information Engineering, Baise University, Baise 533000, China}

\altaffiltext{1}{E-mail: william\_lo@qq.com (W-J L)}
\altaffiltext{2}{E-mail: yingru\_lin@qq.com (Y-R L)}



\begin{abstract}
In this paper, we report the line-locking phenomenon of the blended narrow absorption lines (NALs) within trough-like broad absorption lines (BALs) in quasar SDSS J021740.96--085447.9 (hereafter J0217--0854). Utilizing the two-epoch spectroscopic observations of J0217--0854 from the Sloan Digital Sky Survey, we find that each of its C\,{\footnotesize IV}, Si\,{\footnotesize IV}, N\,{\footnotesize V}, and Ly$\alpha$ BAL troughs actually contain at least seven NAL systems. By splitting these BAL troughs into multiple NAL systems, we find that the velocity separations between the NAL systems are similar to their doublet splittings, with some of them matching perfectly. Cases like J0217--0854, showing line-locking signatures of NALs within BAL troughs, offer a direct observational evidence for the idea that radiative forces play a significant role in driving BAL (at least for Type N BAL) outflows.

\end{abstract}




\section{Introduction} \label{sec:intro}
{The} line-locking, observational signature of the velocity separation between distinct absorption components, similar to the velocity splitting of an absorption doublet and/or multiplet, was reported as early as the 1970s, not long after the discovery of quasars (e.g., \citealp{Williams1975,Coleman1976,Adams1978,Perry1978}). Since then, the line-locking signatures in narrow absorption lines (NALs; absorption widths of less than 500\,$\rm km\,s^{-1}$) have been largely identified in both broad absorption line (BAL) quasars (absorption widths of at least 2000\,$\rm km\,s^{-1}$) and non-BAL quasars (e.g., \citealp{Foltz1987,Borra1996,Tripp1997,Srianand2000,SrianandandPetitjean2000,Vilkoviskij2001,Srianand2002,
Ganguly2003,Fechner2004,Gallagher2004,Benn2005,Misawa2007,Misawa2013,Misawa2014, Misawa2016,Misawa2018,Simon2010,Hamann2011,Bowler2014}). Of particular note is a systematic study by \citet{Bowler2014} that showed evidence for line-locking signatures in almost two-thirds of quasars possesses multiple C\,{\footnotesize IV} absorption systems, based on a sample of about 34,000 quasar spectra \citep{Schneider2010} from the Sloan Digital Sky Survey (SDSS; \citealp{York2000}). This study indicates that line-locked C\,{\footnotesize IV} doublets are a very common feature of NAL outflows. Previous studies have also shown that the probability of a line-locking signature accidentally occurring over a relatively small redshift path is negligible \citep{Foltz1987,Srianand2000,Srianand2002,Ganguly2003,Benn2005}, and line-locking is usually interpreted as an observational signature of radiative acceleration (e.g., \citealp{Mushotzky1972,Scargle1973,Braun1989}). Therefore, the line-locking signature can serve as a reliable criterion to distinguish intrinsic NALs from intervening NALs (e.g., \citealp{Misawa2018}, and references therein). 

\begin{figure*}
\includegraphics[width=2.1\columnwidth]{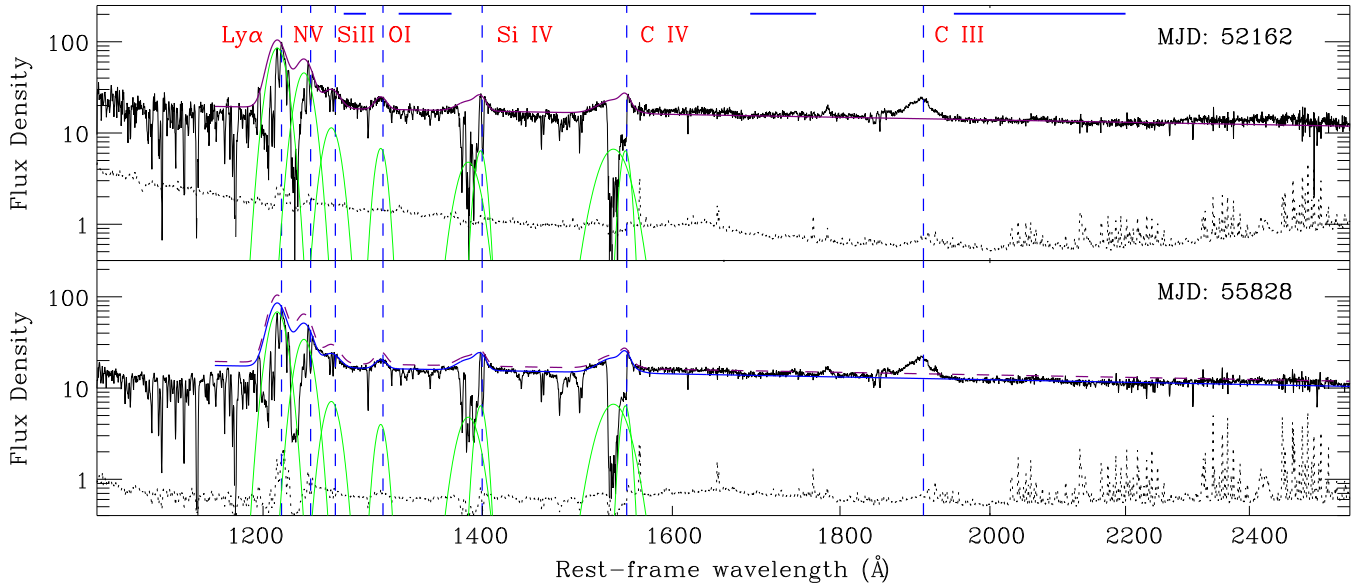}
\caption{Spectra of quasar J0217--0854. The SDSS MJDs of the spectra are labeled. The flux density is in units of $\rm 10^{-17}~erg~s^{-1}~cm^{-2}$. The blue vertical dashed lines mark out the main emission lines. The blue horizontal bars {on top of} the spectra are the regions used to fit the power-law continua. The purple and blue solid lines along the two spectra profiles are the pseudo-continuum fits for the spectra. The purple dashed line in the lower panel is the final pseudo-continuum fit for the MJD 52162 spectrum. The dotted lines near the bottom of each panel are the formal 1$\sigma$ errors. Fits to the emission lines are given as green Gaussian profiles in the bottom of each panel. Both the longitudinal and transverse axes are logarithmic.}  \label{fig.1}
\end{figure*}
Though the radiative line-driving in NAL outflows has been proven through confirming the line-locking phenomenon of intrinsic NALs in both BAL and non-BAL quasars, there is no convincing evidence to confirm the line-driven radiative BAL outflows. One piece of  evidences for line-driven radiative BAL outflows {is the double} trough signature of Ly$\alpha$--N\,{\footnotesize V} line-locking (the so-called ``ghost of Ly$\alpha$") seen in the mean profile of BALs \citep{Weymann1991,Korista1993,Arav1994, Arav1995,Arav1996,North2006,Cottis2010}. \citet{Arav1995} suggested that the Ly$\alpha$--N\,{\footnotesize V} line-locking signature was due to the increased radiation pressure of Ly$\alpha$ photons on outflowing N\,{\footnotesize V} ions, which can serve as evidence for radiative acceleration in quasars {(see also  \citealp{Arav1994,Arav1996})}. However, observational evidence based on a large sample of SDSS quasars \citep{Schneider2010} proved that the presence of the ``ghost of Ly$\alpha$" in the objects that meet all five physical criteria suggested by \citet{Arav1996} is elusive \citep{Cottis2010}.

\defcitealias{Lu2018b}{Paper~I}
Recently, Lu \& Lin (\citeyear{Lu2018b}, hereafter \citetalias{Lu2018b}) have found that some BALs (hereafter Type N BALs\footnote{It is the same as the ``Type II BALs" in Paper I. It is now changed in order to avoid confusion with the narrow-line AGN that exhibits BAL
troughs. Corresponding to the Type N BALs, we rename the other relatively smooth BAL troughs that cannot be decomposed into multiple NALs as ``Type S BALs” (which were called as ``Type I BALs" in Paper I).}) are complexes of NALs in reality and that the splitting of a Type N BAL into multiple NALs can serve as a useful way for probing quasar outflows. In this paper, we report the discovery of the line-locking signatures of NALs within BALs in quasar SDSS J021740.96--085447.9 (hereafter J0217--0854) as a warm-up of our systematic study program on Type N BALs (Lu, W.-J. et al., in preparation). This interesting discovery may offer a convincing evidence for line-driven radiative BAL outflows. The paper is structured as follows. Section \ref{sec:spectro} presents the quasar spectra and describes how we identified the NALs and their line-locking signatures. Section \ref{sec:disscu} contains a discussion and a brief conclusion. Throughout this paper, a $\Lambda$CDM cosmology with parameters $H_0=70\,\rm km\,s^{-1}\,Mpc^{-1}$, $\Omega_{\rm M}=0.3$, and $\Omega_{\Lambda}=0.7$ \citep{Spergel2003} is adopted.
 
\begin{figure}
\includegraphics[width=1.01\columnwidth]{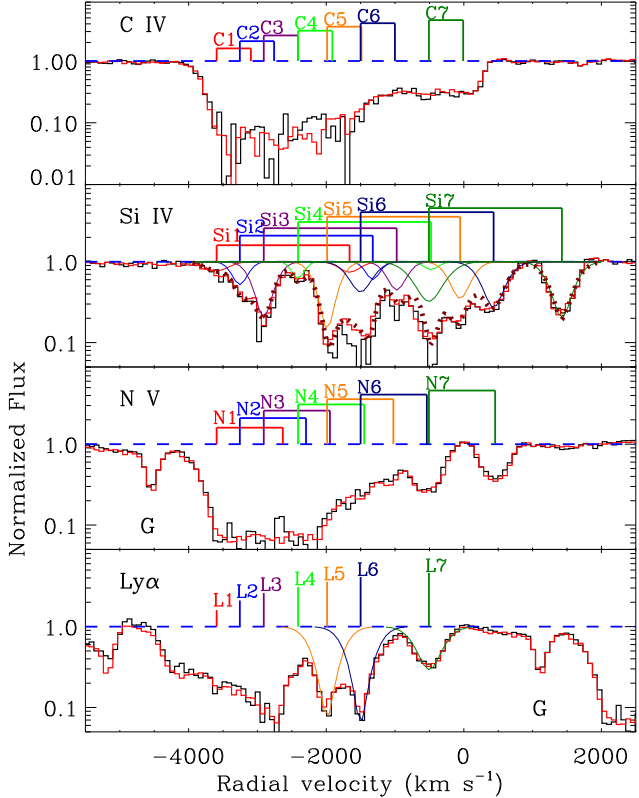}
\caption{Portions of the normalized spectra of J0217--0854, showing the NAL systems detected within the BALs in different ions. The black and red lines show the normalized spectra from observations on MJD 52162 and 55828, respectively. Normalized fluxes are plotted  versus the radial velocity with respect to an emission-line redshift of 2.568. The red, blue, purple, green, orange, dark blue, and dark green vertical lines in each panel mark out the {seven} identified NAL systems. {Gaussian fits with corresponding colors were applied to the NALs within Si\,{\footnotesize IV} and Ly$\alpha$ BALs.} The brown dotted lines in the Si\,{\footnotesize IV} panel represent the total fit model. ``G" marks the strong Si\,{\footnotesize II} $\lambda$1260 Galactic absorption line at redshift of 2.455. The vertical axis is logarithmic.
\label{fig.2}}
\end{figure}

\section{SPECTROSCOPIC ANALYSIS} \label{sec:spectro} 

\subsection{{Continuum fits}} \label{sec:continuum} 
The first spectrum of J0217--0854 {($z=2.568$; \citealp{Paris2017})} is from SDSS, observed on MJD 52162, and the second spectrum is originally from the Baryon Oscillation Spectroscopic Survey (\citealp{Dawson2013}) and was taken on MJD 55828. The MJD 52162 and MJD 55828 spectra have wavelength coverages of $\sim$3800--9200\,\AA~and $\sim$3600--10000\,\AA, respectively. Both of these two spectra have resolutions $\sim$1500 at 3800\,\AA~and $\sim$2500 at 9000\,\AA. The MJD 52162 spectrum has a median signal-to-noise ratio (S/N) of 16.99  per pixel, while the MJD 55828 spectrum has an S/N of 19.93 per pixel. 

As a BAL quasar, J0217--0854 has been studied in many systematic BAL studies \citep{Trump2006,Gibson2009a,Scaringi2009,Allen2011,Filiz2013,Wang2015,He2017}. Its SDSS spectrum has the balnicity indices (BIs; adopting the definition of \citealp{Weymann1991}) of 390.2 and 265.8\,$\rm km\,s^{-1}$ for the C\,{\footnotesize IV}  and Si\,{\footnotesize IV} BALs (taken from \citealp{Gibson2009a}), respectively. 

We downloaded the two-epoch spectra of J0217--0854 from SDSS data release 14 (DR14; \citealp{Abolfathi2018}) where the flux calibration has been improved. And then we iteratively fitted the power-law continuum for the spectra based on a few relatively line-free  wavelength regions (1270--1290, 1320--1370, 1690--1770, and 1950--2200\,\AA~in the rest frame; imitating the definition of \citealp{Gibson2009a}). To reduce the influences from the  remaining sky pixels and/or the emission/absorption lines on our fittings, we left out the pixels that {fall} beyond the formal 3$\sigma$ errors level from the fitting model at each iteration (e.g., \citealp{Gibson2009a, Lu2018a}). Besides,  because the BAL systems in J0217--0854 are superimposed on the emission lines, appropriate Gaussian fits should be applied to the emission lines. Specifically, each of the C\,{\footnotesize IV} $\lambda$1549 and Si\,{\footnotesize IV} $\lambda$1400 emission line was fitted with two Gaussians, while each of the O\,{\footnotesize I} $\lambda$1305, Si\,{\footnotesize II} $\lambda$1262, N\,{\footnotesize V} $\lambda$1240, and Ly$\alpha$ $\lambda$1215 emission lines was fitted with a single Gaussian, respectively. We attribute no physical meaning to these Gaussian profiles, which were simply used to model the emission-line profiles in order to properly identify the absorption lines. Combining the power-law continuum fit and the fits to the emission lines, we got the pseudo-continuum fit for each spectrum (see Figure \ref{fig.1}).

\begin{table}
    \centering
\caption{Velocity Splitting among Distinct Components\label{tab.1}}
\begin{tabular}{lcrcccrc} 
\hline 
\hline 
Doublet &\multicolumn3c{$\rm Splitting^a$} & Components & \multicolumn3c{$\rm Separation^b$} \\
             &\multicolumn3c{($\rm km\,s^{-1}$)} & &\multicolumn3c{($\rm km\,s^{-1}$)} \\
\hline
C\,{\footnotesize IV}&&499&&3-4&&499&\\
                                         &&&&5-6&&487&\\                                           
Si\,{\footnotesize IV}&&1933&&4-7&&1902&\\                                              
N\,{\footnotesize V}&&962&&3-5&&922&\\
                                          &&&&4-6&&909&\\
                                          &&&&6-7&&993&\\
\hline 
\end{tabular}
\begin{tablenotes}
\footnotesize
\item{$^{\rm a}$The laboratory value.}
\item{$^{\rm b}$The measured separation in the spectra.}
\end{tablenotes}
\end{table}

\subsection{{Identification of NALs}} \label{sec:NALs} 
We then tried to identify NALs within the C\,{\footnotesize IV}, Si\,{\footnotesize IV}, N\,{\footnotesize V}, and Ly$\alpha$ BAL troughs in the spectra normalized by the pseudo-continua. We show these absorption lines as the normalized fluxes versus the radial velocity {of the strongest member, for each absorption doublets,\footnote{{Namely, 1548.195, 1393.755, 1242.804, and 1215.670\,\AA~for C\,{\scriptsize IV}, Si\,{\scriptsize IV}, N\,{\scriptsize V} absorption doublets, and the Ly$\alpha$ absorption line, respectively \citep{Verner1994}.}} with respect to the redshift of the emission line of 2.568} (see Figure \ref{fig.2}). As shown in Figure \ref{fig.2}, the NALs for each ion are severely blended, resulting in trough-like BAL profiles {in C\,{\footnotesize IV} and N\,{\footnotesize V}. This is consistent with the previous findings that saturation is common in BALs (e.g., \citealp{Arav1999}; \citealp{Lu2018c}). It requires skill to identify NALs within {BAL troughs}. Usually, }they can be split by using absorption together lines from different ions.  For instance, the Si\,{\footnotesize IV} $\lambda\lambda$1394, 1403 absorption doublet is usually easier to be identified because it shows weaker EW and its blue and red members separate farther than the C\,{\footnotesize IV} $\lambda\lambda$1548, 1551 and N\,{\footnotesize V} $\lambda\lambda$1239, 1243 absorption doublets, {owning to} the differences in oscillator strength and fine structure between these three ions ({see \citetalias{Lu2018b}}). 
In addition, blended NALs for other ions can be split according to the Ly$\alpha$ absorption, because its single absorption line structure can result in less blended profiles {\citep{Verner1994}.

In J0217--0854, the kinematic components were identified based on the visible NAL profiles within the Si\,{\footnotesize IV} and  Ly$\alpha$ BALs, and seven NAL systems within each BAL trough were identified (Figure \ref{fig.2}). 
{Here, we elaborate on the steps we have taken to identify the NALs. 
First, we identified the NAL systems $5\thicksim7$ based on the visible NAL profiles within the Ly$\alpha$ BALs. To determine more correctly the line centers of the NAL systems, Gaussian fits were applied to these visible NAL profiles (L5$\thicksim$L7, see Figure \ref{fig.2}). 
Second, we identified the NAL systems $1\thicksim4$ based on the blue components of Si\,{\footnotesize IV} doublets (since their red components are blended with the blue components of systems $5\thicksim7$).
Third, following \citetalias{Lu2018b}, we employed seven pairs of Gaussian functions to fit the blended NALs within the Si\,{\footnotesize IV} BAL trough in the second-epoch spectrum. During the fitting, the velocity separation of each doublet is fixed to the laboratory value. The line centers of systems $5\thicksim7$ are fixed to the values obtained in the first step, while the line centers of systems $1\thicksim4$ are allowed to be fine-tuned. 
Finally, we marked out the identified NAL systems in C\,{\footnotesize IV} and N\,{\footnotesize V} BALs. {The} selection of doublet wavelengths is performed by anchoring the positions of all blue components with the corresponding resolved Si\,{\footnotesize IV} blue components and Ly$\alpha$ troughs.}

As shown in Figure \ref{fig.2}, in velocity space, some of the narrow absorption components within BALs are locked together with their doublet splittings. Because the pixel width of SDSS spectra is 69\,$\rm km\,s^{-1}$, we search for line-locked NAL pairs within an error of 69\,$\rm km\,s^{-1}$ too (e.g., \citealp{Bowler2014}). Note that although all the NAL systems in the C\,{\footnotesize IV} BAL trough (as well as in most of the N\,{\footnotesize V} BAL trough) are entirely blended, their velocities are confirmed by our fits of the Si\,{\footnotesize IV} and Ly$\alpha$ absorption lines. Thus, even the C\,{\footnotesize IV} doublets inside broad trough show no evidence of structured absorption, their line-locking signatures are identifiable. Finally, we discovered two line-locked NAL pairs with a velocity splitting corresponding to the C\,{\footnotesize IV} doublet splitting, one line-locked NAL pair with a velocity splitting corresponding to the Si\,{\footnotesize IV} doublet splitting and three line-locked NAL pairs with a velocity splitting corresponding to the N\,{\footnotesize V} doublet splitting (see Table \ref{tab.1} for details). 

\section{DISCUSSION AND CONCLUSION} \label{sec:disscu}
The results in Section \ref{sec:spectro} provide important information on the quasar outflows. First, the phenomenon of multiple line-locked NALs within a BAL trough strongly favors the ideas that the BAL outflow (at least in J0217--0854) is made up of several physically separated clumpy structures with similar locations, kinematics, and physical conditions and that the radiative line-driving plays an important role in the acceleration of these clumpy clouds. The line-locking signatures in J0217--0854 also indicate that our line of sight is approximately parallel to the wind streamlines (e.g., \citealp{Hamann2011}). We can make a rough estimate of the inclination of the wind streamlines with respect to our line of sight based on the Si4--Si7 line-locked NAL pair (we choose this line-locked pair because it retains relatively complete Gaussian profiles, which are useful for more correctly estimating the outflow velocities; see Figure \ref{fig.2}). The velocity-split difference with respect to the laboratory splitting of the Si4--Si7 pair is $\thicksim31\,\rm\,km\,s^{-1}$ (see Table \ref{tab.1} and Figure \ref{fig.2}), which implies the outflow trajectory is $\thicksim10^\circ$ of the line of sight.

Second, anticorrelation between the fractional variation of the ionizing continuum and ultraviolet (UV) outflow lines has recently been proved (\citealp{Lu2017,Lu2018a}), revealing the ubiquitous effect of the ionizing continuum variability on the UV outflow lines. In the case of J0217--0854, although the pseudo-continuum shows a fractional weakening of $0.129\pm0.029$ between the two observations (Figure \ref{fig.1}), the BALs show no significant {($\Delta$EW\textless1$\sigma^{'}_{\rm EW}$)\footnote{{The definition is  $\rm \frac{\Delta EW}{\sigma^{'}_{\rm EW}}=\frac{|EW_2-EW_1|}{\sqrt{\sigma^{2}_{\rm EW_1}+\sigma^{2}_{\rm EW_2}}}$, where $\rm EW_1$ and $\rm EW_2$ are EWs of a BAL for two epochs and $\sigma^{2}_{\rm EW_1}$ and $\sigma^{2}_{\rm EW_2}$ are their uncertainties. EW and $\sigma_{\rm EW}$ are calculated using equations (2) and (3) in \citetalias{Lu2018b}.}}} variability on rest-frame timescales of about 2.8 years. The lack of variability in these outflow lines is probably due to their line saturation \citep{Lu2018c}. However, in this case, the line-locking signatures {are} still visible, indicating that the line-locking signature can remain steady for at least a few years (e.g., \citealp{Vilkoviskij2001}).

Third, the absorption depth of C\,{\footnotesize IV} and Si\,{\footnotesize IV} BAL troughs of J0217--0854 is deeper than their corresponding broad emission lines (Figure \ref{fig.1}), which indicates that the absorption regions should be extended enough to occult the broad emission line region (BELR). Besides, it can be inferred from the C\,{\footnotesize IV} BAL that the NAL systems $1\thicksim4$ are very close to zero intensity while the NAL systems $5\thicksim7$ {show residual light at their bottom}, suggesting that the systems $1\thicksim4$ almost completely occult the BELR while the systems $5\thicksim7$ only partially cover the BELR. Such a situation, the complete and partial coverage {coexisting} in a single BAL system, is reported for the first time {in the framework of BALs formed by blended NAL complexes}.

Finally, as pointed out by \citet{Hamann2011}, it is still unclear whether radiation pressure can actually lock the UV absorption doublets together and, if this phenomenon is really feasible, then how the line-locking is performed in quasar outflows. Our discovery, the multiple line-locked NALs within BALs, may offer more observational materials on the study of these problems. Systematic investigation of the line-locking signatures in Type N BALs will be presented in the future work.

\acknowledgments

We gratefully thank the anonymous referee for helpful comments, which have significantly improved both the scientific content and the general appearance of our manuscript.  We thank Yiping Qin for helpful discussions.

Funding for SDSS-III was provided by the Alfred P. Sloan Foundation, the Participating Institutions, the National Science Foundation, and the U.S. Department of Energy Office of Science. The SDSS-III website is \url{http://www.sdss3.org/}.

SDSS-III is managed by the Astrophysical Research Consortium for the Participating Institutions of the SDSS-III Collaboration, including the University of Arizona, the Brazilian Participation Group, Brook haven National Laboratory, Carnegie Mellon University, University of Florida,the French Participation Group, the German Participation Group, Harvard University, the Instituto de Astrofisica deCanarias, the Michigan State/Notre Dame/JINA Participation Group, Johns Hopkins University, Lawrence Berkeley National Laboratory, Max Planck Institute for Astrophysics, Max Planck Institute for Extraterrestrial Physics, New Mexico State University, New York University, Ohio State University, Pennsylvania State University, University of Portsmouth, Princeton University, the Spanish Participation Group, University of Tokyo, University of Utah,Vanderbilt University, University of Virginia, University of Washington, and Yale University.

\bibliographystyle{aasjournal}
\bibliography{NALvsBAL} 

\end{document}